\documentclass[preprint,tightenlines,superscriptaddress,
prd,nofootinbib,eqsecnum,showpacs]{revtex4}

\usepackage{amssymb}
\def\dd{{\rm d}}

\def\_#1{^{}_{#1}}
\def\half{{\textstyle\frac12}}

\def\beq{\begin{equation}}
\def\eeq{\end{equation}}
\def\bea{\begin{eqnarray}}
\def\eea{\end{eqnarray}}
\usepackage{enumerate}

\begin{document}

\title{Effects of Dynamical Compactification on $D$-Dimensional Gauss-Bonnet FRW Cosmology}

\author{Keith Andrew}\email{keith.andrew@wku.edu}
\affiliation{Department of Physics and Astronomy, Western Kentucky University\\
Bowling Green, KY 42101, U.S.A.}

\author{Brett Bolen}\email{brett.bolen@wku.edu}
\affiliation{Department of Physics and Astronomy, Western Kentucky University\\
Bowling Green, KY 42101, U.S.A.}

\author{Chad A. Middleton}\email{chmiddle@mesastate.edu}
\affiliation{Department of Physical and Environmental Sciences, Mesa
State College, Grand Junction, CO 81501, U.S.A.}


\vspace{2cm}

\begin{abstract}
We examine the effect on cosmological evolution of adding a string
motivated Gauss-Bonnet term to the traditional Einstein-Hilbert
action for a $(1+3) + d$ dimensional Friedman-Robertson-Walker (FRW)
metric. By assuming that the additional dimensions compactify as the
usual $3$ spatial dimensions expand, we find that the Gauss Bonnet
terms give perturbative corrections to the FRW equations. We find
corrections that appear in the calculation of both the Hubble
constant, $H_0$, and the acceleration parameter, $q_0$, for a variety of
cases that are consistent with a dark energy equation of state.
\end{abstract} \maketitle
\section{Introduction}

During the last decade observational cosmology has advanced to the
point of placing a number of
 constraints on our theoretical understanding of the universe.
 Observations of Type Ia Supernovae (SNe Ia) have become a critical probe of the
  expansion history of the universe \cite{experiment1} and recent measurements of
  the cosmic microwave background indicate an accelerated expansion of the universe occurred in an
  early inflationary epoch \cite{experiment2}. Current efforts to
  understand these observations within a mathematically consistent framework often
   start with the ten or eleven dimensional spacetime manifolds motivated
  by superstring or M theory \cite{Witten} \cite{Ohta}. These theories often
   require a compactification phase to achieve the four dimensional spacetime manifold
    associated with the current universe and a means for describing the six
     dimensional manifold in terms of current observational constraints, i.e. they are not obviously visible. The
  observations indicate that the expansion of the universe is accelerating,
   at critical density, has a nonzero cosmological constant and is flat. These
  observations also estimate that some 70\% of the universe is composed of
  dark energy, another 25\% dark matter and the rest baryonic matter.
  The dark energy is often characterized by a negative pressure equation
   of state $p=w \rho$ where the equation of state parameter for the exotic matter, $w$, lies near
  -1. One candidate for the exotic matter is a type of dark energy associated with a cosmological constant arising from
  zero point vacuum energy fluctuations \cite{Milton} \cite{Kantowski}that give a constant value of $ w=-1$ or quintessence arising
  from a time varying scalar field which gives a dark energy
   that decreases with time $ -1 < w < -1/3 $ where dark energy density decreases with scale factor a(t) as $ \rho \propto a^{(-3(1+w))}$
   or phantom energy corresponding to cases of exotic matter with $ w < -1 $ \cite{Peebles} \cite{Caldwell} .

An understanding of dark energy can be searched for in
string/M-theory and the resulting generalizations of low energy
gravity that are motivated by such investigations. It is possible to
utilize unusual couplings to matter fields \cite{randall} or a
dilaton field \cite{fang}, or coupling constants \cite{Vereshchagin}
extra dimensions \cite{Zhu} or Kaluza Klein compactifications
\cite{Mohammedi}, or a time varying cosmological constant
\cite{sussman} or dimensional reductions \cite{Filippov} or
corrections to the pure gravity sector of the theory through
interesting functions of the curvature \cite{weinberg} or
Gauss-Bonnet motivated couplings \cite{Calcagni}or couplings to
branes \cite{Padilla}. The use of phantom fluids where the pressure
and energy density are related by an equation of state given by
$p<-\rho$; which for a scalar field can be achieved by using a
negative kinetic energy term in the Lagrangian density, is another
promising avenue of development \cite{caldwell}\cite{Kujat}. One
demands that such theories give reasonable descriptions of the large
scale four dimensional universe described by the
Friedmann-Roberston-Walker (FRW) equations arrived at from the
standard Einstein field equations. An insightful way to approach the
dark energy problem is to utilize modifications of the Einstein
field equations; to search for ways to accommodate the dark energy
observations within the context of higher dimensional or higher
order modifications to the theory or to use noncommutive geometries
and Casimir \cite{Fabi} energy or by equivalent scalar-tensor
theories or moduli fields \cite{Cavaglia}. Nojiri and Carrol examine
positive and negative powers of the scalar curvature in order to
describe a stable inflationary phase and an accelerating late phase
in geometric terms \cite{Nojiri}\cite{Sotiriou}\cite{Carrol}. Within
the framework of General Relativity the cosmic acceleration can be
accommodated for with a phantom equation of state. Lovelock
modifications consist of dimensionally extended Euler densities
which include higher order curvature invariants in such a way that
they lead to second order gravitational field equations but are
always total derivatives in four dimensions \cite{Zumino}. The
quadratic curvature terms are referred to as Gauss-Bonnet terms
which come about in the low energy limit of heterotic string theory
\cite{Gross}\cite{deser}.

Here we consider a dynamic compactification of a $(4+d)$-dimensional
manifold to a maximally symmetric manifold of dimension $d$ and an
expanding FRW space of dimension $4$. We modify the gravity
Lagrangain to first order in string tension $\alpha$ by a
Gauss-Bonnet term that is quadratic in the curvature\cite{Nojiri2}.
Here we have considered these terms explicitly coupled to matter
fields obeying some equation of state with the strength of each term
treated as a parameter to be fixed by observations. It is known that
when the internal space is a nonsingular time independent compact
manifold without boundary then there is no accelerated expansion
unless the strong energy condition is violated and if the higher
dimensional stress energy tensor obeys the strong energy condition
than so does the lower dimensional stress energy tensor \cite{susy}.
Here we will consider the time dependent solutions of arbitrary
dimension on a maximally symmetric manifold thereby avoiding the
time independent nature of the compact space. In this paper we will
investigate the Einstein-Gauss-Bonnet field equations on a manifold
$M^n$ that undergoes dynamical compactification to $M^n = (R^1
\times M^3 )\times S^d$. We will include matter by specifying an
equation of state and applying the constraint equations on the
energy momentum tensor $\partial_\mu T^{\mu \nu}=0$. Within this
model we will extract the Friedmann-Robertson-Walker equations with
the Gauss-Bonnet terms as correction terms to the standard 4-d
equations. These equations are then solved pertubatively and
analyzed within the context of the observational constraints and an
equation of state. To understand the impact of higher order
curvature terms on the observed cosmic evolution we will solve the
Gauss-Bonnet corrected FRW equations that undergo a dynamical
d-dimensional compactification to a maximally symmetric space. This
paper is organized as follows. In Section \ref{framesec} we present
a general Einstein plus Gauss-Bonnet action in $d$-dimensions where
the field equations are calculated and the correction to the FRW
equation is given. In section \ref{dynamiccompact} we prepare a
perturbative solution for special cases including with and without
the cosmological constant and calculate the corrected values of the
Hubble constant and the deceleration parameter. In Section
\ref{conclusion} we summarize our results.

\section{Framework for Field Equations} \label{framesec}
In this paper we will follow the notation of Paul and Mukherjee\cite{PaulMuk} and Mohmmedi
 \cite{Mohammedi} to express the Einstein-Hilbert
action with an extra Gauss-Bonnet term as
\beq\label{action}
     S= \int \dd^D x \sqrt{- g} \left[ R - 2 \lambda - \epsilon \, \mathcal{G} \right]
\eeq where $\mathcal{G}$ is a Gauss-Bonnet term $(\mathcal{G}\equiv
R_{A B C D} R^{A B C D} - 4 R_{A B} R^{A B} +R^2)$. A variation of
the action (\ref{action}) with respect to $g_{AB}$ produces the equation;
\beq\label{field}
 G_A{}^C+\lambda g_A{}^C+\epsilon\;\mathcal{G}_A{}^C =\frac{1}{2\Upsilon}T_A{}^C
 \eeq
where $\Upsilon$ is the $4+d$ dimensional coupling constant and the Einstein and Gauss-Bonnet tensors, respectively, are
\bea
 G_A{}^C&=&R_A{}^C -\half g_A{}^C R \\
 \mathcal{G}_A{}^C&=&\frac{1}{2}\left( R_{B D EF} R^{BDEF} -
4 R_{BD} R^{BD} + R^2\right)  \delta_A{}^C \nonumber\\
&& \quad -\big( 2 R_{BDEA} R^{BDEC} + 2 R R_A{}^C-4 R_D{}^B R_{B A}{}^{D
C}- 4 R_B{}^C R_A{}^B \big) \, .
\eea

Furthermore we will assume that the stress-energy tensor will be that of a perfect fluid,  thus it is of the form
\beq
T_\mu {}^ \nu= \textrm{diag} \left[ \rho(t), p(t) , p(t), p(t), p_d(t), ... p_d(t)\right]
\eeq
where $p_d(t)$ is the pressure of the higher dimensional compact manifold.

We will choose a metric ansatz:
\beq \dd s^2 = - \dd t^2 + a^2(t)
\left[ \frac{\dd r^2}{1- K r^2} + r^2 \left(\dd \theta^2 + \sin^2
\theta \dd \phi^2 \right) \right] + b^2(t)\gamma_{m n} (y) \dd y^m \dd
y^n \eeq where the extra dimension is defined to be maximally
symmetric such that the Riemann tensor for $\gamma$ has the form
$R_{abcd} = \kappa (\gamma_{a c} \gamma_{b d} - \gamma_{a d}
\gamma_{b c})$ . In agreement with current observations we will
consider the usual $3$ spatial dimensions to be flat $(K=0)$ and
also demand that the extra dimensions be flat as well ($\kappa=0$)
in agreement with  the finding that one finds unphysical properties
for $\rho$ and $p$ if $\kappa \ne 0$ \cite{Mohammedi}.

The metric leads to Riemann tensors of the following form (where both the dimensions are flat)
\bea
R_{a 0 a 0} = a \ddot a \, , \quad
R_{a b a b} = a^2 \dot a^2 \, , \quad
R_{m 0 m 0} = b \ddot b \, , \quad
R_{m a m a} =  a \dot a b \dot b \, , \quad
R_{a b a b} =  b^2 \dot b^2
\eea
where $a, b$ are indices which run from $1 ... 3$ and $m,n$ are indices which are in the extra dimensions. (obviously the symmetries as well
$R_{abcd}= R_{badc} =-R_{abdc} =- R_{bacd} $)   The Ricci Tensor and
Ricci Scalar are
\beq
R_{00} =  3 \frac{\ddot a}{a} + d \frac{\ddot b}{b} \, , \quad \quad
R_{a a} =2 \dot a^2 + a \ddot a+d \frac{a \dot a \dot b}{b}  \, , \quad
R_{m m} = 3 \frac{b \dot a \dot b}{a} + (d-1) \dot b^2 + \ddot b b
\eeq
\beq
R = 6 \frac{\ddot a}{a} + 2 d \frac{\ddot b}{b} + 6 d \frac{\dot a \dot b}{a b} + 6\frac{\dot a^2}{a^2}  + d (d-1)\frac{\dot b^2}{b^2}
\eeq
where $d$ is the number of extra dimensions.

The full Einstein tensor may be expressed as having an Einstein term $G_\mu {}^\nu$ and a Gauss-Bonnet term $\mathcal{G}_\mu {}^\nu$.
We make the assumption that the extra dimensions compactify as the $3$ spatial dimensions expand as \cite{Mohammedi}
\beq\label{comp}
b(t) \sim\frac{1}{a^n(t)}
\eeq
 where $n>0$ in order to insure that the scale factor of the compact manifold is both dynamical and compactifies as a function of time.
 With this ansatz, the non-zero elements of the Einstein tensor takes the form
\bea
G_0 {}^0 &=&\frac{1}{2}\left[6+d \, n(d \, n-n-6)\right]\frac{\dot a^2}{a^2}\nonumber\\
&=& \eta_1 \, \frac{\dot a^2}{a^2} \label{E00} \\
G_a{}^a&=&\left(d n-2 \right)\frac{\ddot a}{a} -\frac{1}{2} \left[2+d n(d n +n-2)\right] \frac{\dot a^2}{a^2}
 \nonumber \\
  &=& \eta_2 \,\frac{\ddot a}{a}+ \eta_3 \, \frac{\dot a^2}{a^2} \label{Eaa} \\
G_m{}^m &=&\left(d n-  n-3  \right) \frac{\ddot a}{a}-\frac{1}{2} \left[6+n(d-1)(d \, n-4)\right]\frac{\dot a^2}{a^2} \nonumber\\
  &=& \eta_4 \, \frac{\ddot a}{a}+ \, \eta_5 \, \frac{\dot a^2}{a^2}
  \label{Emm} \, .
\eea
Note that when $d=4$ the equations become the well known FRW equations in four dimensions.
In $4+d$ dimensions, the Gauss-Bonnet terms become
\bea
\mathcal{G}_0 {}^0 &=& d \, n \left[(d-1)n\left[\frac{1}{2}(d-2)n\left[(d-3)n-12\right]+18\right]-12\right]\frac{\dot a^4}{a^4}\nonumber\\
&=& \xi_1 \frac{\dot a^4}{a^4}\\
\mathcal{G}_a {}^a &=& \frac{1}{2}d \, n\Bigg[ (d+1)(d-1) (d-2) n^3+4 (2-d (d+1))
  n^2-4(d-3)n+8 \Bigg]\frac{\dot a^4}{a^4}\nonumber\\
&&\quad - 2d \, n\Bigg[ (d-1)n\left[ (d-2)n-6 \right]+6\Bigg]\frac{\ddot a\dot a^2}{a^3} \nonumber\\
&=& \xi_2 \,\frac{\dot a^4}{a^4} + \xi_3 \, \frac{\ddot a\dot a^2}{a^3} \\
\mathcal{G}_m {}^m &=& \frac{1}{2}(d-1)d \, n^2 \Bigg[(d-2)n\left[(d-3)n-8\right]+12\Bigg] \frac{\dot a^4}{a^4} \nonumber\\
& &\quad- \Bigg[2(d-1)n\Bigg[(d-2)n\left[(d-3)n-9\right]+18\Bigg]-12\Bigg]\frac{\ddot a\dot a^2}{a^3} \nonumber\\
&=& \xi_4 \, \frac{\dot a^4}{a^4}  + \xi_5 \, \frac{\ddot a \dot a^2}{a^3}
\eea
where the constants $\eta_i$ and $\xi_i$ depend upon the values of $n$ and $d$ as defined
above. Note that if we allow $d \rightarrow 0$ then the Gauss-Bonnet terms vanish for $\mathcal{G}_0 {}^0$
and $\mathcal{G}_a {}^a$ as one would expect in four dimensions.

\section{Dynamic Compactification} \label{dynamiccompact}
 For the case of both spaces having flat curvature, the
$D$-dimensional  Friedmann-Robertson-Walker (FRW) equations (eqn \ref{field}) take the form
\bea
\frac{\rho}{2 \Upsilon} &=&\frac{1}{3} \eta_1 \left[ 3\frac{\dot a^2}{a^2} - 3\frac{\lambda}{\eta_1} \right] + \epsilon \, \xi_1 \, \frac{\dot a^4}{a^4}  \label{density1}\\
\frac{p}{2 \Upsilon} &=& \left[\eta_2 \, \frac{\ddot a}{a} + \eta_3 \, \frac{\dot a^2}{a^2} + \lambda \right] + \epsilon \left( \xi_2 \frac{\dot a^4}{a^4}  + \xi_3 \, \frac{\ddot a\dot a^2}{a^3}\right)  \label{presure1} \\
\frac{p_d}{2 \Upsilon} &=&\left[ \eta_4 \, \frac{\ddot a}{a} + \eta_5 \frac{\dot a^2}{a^2} +\lambda \right]+ \epsilon \left( \xi_4 \frac{\dot a^4}{a^4}
 + \xi_5 \, \frac{\ddot a\dot a^2}{a^3}\right)\,\label{presD}
\eea
 As
the scale factor of the compact manifold tends to zero with time, we
will make the further assumption that the pressure in the extra $d$
dimensions is zero $(p_d=0)$. Together with these Einstein
equations, we also demand that the conservation equation hold for
the stress-energy tensor $\left( \nabla_\mu T^\mu{}_\nu=0 \right)$
or
\beq\label{cons}
  \left\{\frac{\dd}{\dd t} (a^3 \rho) + p \, \frac{\dd}{\dd t} (a^3) \right\} + d \, a^3 \, \frac{\dot b}{b} \left( \rho +  p_d \right) = 0
\eeq
Using the assumption that
$b=1/a^n$, this becomes
\beq \frac{\dd}{\dd t} (a^3 \rho) +
\tilde{p} \, \frac{\dd}{\dd t} (a^3)= 0 \eeq which by simple algebra
may be written in the more familiar form \beq\label{EoS}
\dot{\rho}+3\frac{\dot{a}}{a}(\rho+\tilde{p})=0 \, . \eeq Note that
this is simply a statement that $\dd E = - P \, \dd V$ where we have
defined an ``effective" pressure $\left( \tilde p \right)$
 \cite{Mohammedi} which an observer
constrained to exist only upon the ``usual" $3$ spatial dimensions
would see as \beq\label{effpres}
\tilde{p}=p-\frac{1}{3}d\,n\;(\rho+p_d). \eeq This effective
pressure can be easily computed from the $d$-dimensional FRW
equations (\ref{density1})-(\ref{presD}) and is given by the
relation \bea\label{presT} \frac{\tilde{p}}{2
\Upsilon}&=&\frac{1}{3} \eta_1\left[- \left(2 \frac{\ddot a}{a} +
\frac{\dot a^2}{a^2}\right) +3\frac{\lambda}{\eta_1} \right]+
\epsilon \left(  \gamma_1 \, \frac{\ddot a\dot a^2}{a^3}+\gamma_2
\frac{\dot a^4}{a^4} \right) \eea where we've defined the constants
\beq
\gamma_1=\xi_3-\frac{1}{3}dn\;\xi_5\;\;,\;\;\gamma_2=\xi_2-\frac{1}{3}dn\;(\xi_1+\xi_4).
\eeq
 Note that the above field equations (\ref{density1})-(\ref{presD}), (\ref{presT}) become the same as Mohammedi \cite{Mohammedi} in the limit where
$\epsilon \rightarrow 0$ (ie. No Gauss-Bonnet terms).  By redefining the coupling and cosmological constant in eqns. (\ref{density1}) and (\ref{presT}), one recovers standard 4-D FRW cosmology for arbitrary values of $n$ and $d$ in the $\epsilon \rightarrow 0$ limit.

\subsection{Asymptotic solution when $\lambda=0$} \label{lambda0}

We will first examine the case when the Gauss-Bonnet terms are small with respect to the Einstein terms so they may be treated as perturbations.
We take the pressure in the extra dimensions, $p_d$, to be zero so  equation (\ref{presD}) becomes
\beq\label{adot}
\frac{p_d}{2 \Upsilon} = 0= \eta_4 \frac{\ddot a}{a} +\eta_5 \frac{\dot a^2}{a^2} + \epsilon \left( \xi_4 \frac{\dot a^4}{a^4} + \xi_5 \frac{\ddot a}{a}\frac{\dot a^2}{a^2}\right)
\eeq
This equation may be solved perturbatively by assuming that $a (t) \simeq a_0 (t) +\epsilon \;a_1 (t)$.     Collecting terms in
$\epsilon$, we get a zeroth order equation
\beq
\frac{\ddot a_0}{a_0} +\frac{\eta_5}{\eta_4}\frac{ \dot a_0^2}{a_0^2} =0.
\eeq
This equation yields the solution
\beq
a_0(t) = (\mu t + \nu)^{\eta_4/(\eta_4+\eta_5)}
\eeq
 where $\mu$ and $\nu$ are constants of integration and $\eta_4+\eta_5\neq 0$.\footnote{ $\eta_4+\eta_5= 0$ demands a complex value of $n$ which will not be considered here.}  This solution can be used to find an expression for the  zeroth-order energy density and effective pressure.  We find
\bea
\frac{\rho_0}{2 \Upsilon}& &=\mu^2\frac{\eta_1\eta_4^2}{(\eta_4+\eta_5)^2}\cdot\frac{1}{a_0^{2(\eta_4+\eta_5)/\eta_4}}\label{rho0}\\
\frac{\tilde{p_0}}{2 \Upsilon}& &=\frac{1}{2}\mu^2\alpha_1\frac{\eta_4\left(\eta_4-2\eta_5\right)}{(\eta_4+\eta_5)^2}\cdot\frac{1}{a_0^{2(\eta_4+\eta_5)/\eta_4}}\label{tildep0}
\eea Combining (\ref{rho0}) and (\ref{tildep0}), we can eliminate
$a_0$ and obtain an equation relating $\tilde{p}$ and $\rho$ of the
form \beq\label{EoSl0}
\tilde{p}_0=-\frac{1}{3}\left(1-2\frac{\eta_5}{\eta_4}\right)\rho_0
= w \rho_0 . \eeq
This equation gives a constraint on the possible
values of $n$ and $d$ by choosing a value for $w$ and so it mimics
an equation of state (EOS).  At first glace it seems odd that the
dynamical equations themselves impose a relation between $p$ and
$\rho$, but this happened because of our choice that the extra
dimensions compactify as  $b(t) \sim 1/a(t)^n$.  Equation
\ref{EoSl0} may be solved for $w$ as
 \beq w = \frac{1}{3} \left(
\frac{3+(d-1)(d \, n -3)n}{n(d-1)+3}\right) \, . \eeq Note, in the
limit where $n \rightarrow 0$, $w=1/3$ which is the relationship one
would expect for a radiation dominated universe. Thus, the
geometrical terms in the compactifacation are playing the same role
as matter.  Thus, by demanding that $w$ have a physical value; one
may use this relationship to restrict the choices of $n$ and $d$.
For instance if $d=7$, then $n$ must be less then $1/2$ if $w$ is
demanded to have a physically reasonable value of between $1$ and
$-2$.

The first order perturbation is of the form
\beq
 \ddot a_1(t) +\left(2\frac{\eta_5}{\eta_4}\frac{ \dot a_0}{a_0}\right)\dot{a}_1(t)+\left(3 \frac{\ddot a_0}{a_0}
            + 2\frac{\eta_5}{\eta_4} \frac{\dot a_0^2}{a_0^2}\right)a_1(t)=
            -\frac{1}{\eta_4}\left(\xi_4 \,\frac{ \dot a_0^4}{a_0^4}+ \xi_5\frac{ \ddot a_0}{a_0}\frac{ \dot a_0^2}{a_0^2}\right) \, a_0
\eeq Using the solution for $a_0$ as found above, the solution for
$a_1$ can be found and is of the form \beq
a_1(t)=a_{1,\;hom}(t)+a_{1,\, non}(t) \, . \eeq The homogeneous
solution is given by \beq\label{a1h} a_{1,hom}(t)=A(\mu
t+\nu)^{\eta_4/(\eta_4+\eta_5)}+B(\mu
t+\nu)^{-\eta_5/(\eta_4+\eta_5)} \eeq where $A$ and $B$ are
constants of integration and $a_{1 \, non}$ is the non-homogeneous
solution. The resulting non-homogeneous differential equation is of
the Cauchy-Euler form (if you plug in for $a_0$): \beq \ddot a_{1,
non}(t) - \frac{2\mu(\beta-1)}{(\mu t+\nu)} \, \dot a_{1, non}
(t)+\frac{\mu^2 \beta(\beta-1)}{(\mu t+\nu)^2} \, a_{1, non}(t) =
\frac{\beta^4\xi_5}{\eta_4}\left[\frac{\eta_5}{\eta_4}-\frac{\xi_4}{\xi_5}\right]\frac{\mu^4}{(\mu
t+\nu)^{4-\beta}} \eeq where we've defined the quantity \beq \beta =
\frac{\eta_4}{(\eta_4+\eta_5)} \eeq The non-homogeneous solution is
$a_{1, non}(t)=C \, (\mu t+\nu)^{\beta-2}$ where \beq C=\frac{\mu^2
\beta^4\xi_5}{2\eta_4}\left[\frac{\eta_5}{\eta_4}-\frac{\xi_4}{\xi_5}\right].
\eeq Thus the general solution for $a(t)$ to first order is \beq
a(t) =(\mu t+\nu)^\beta\left[1+\epsilon\left[A+\frac{B}{(\mu
t+\nu)}+\frac{C}{(\mu t+\nu)^2}\right]\right]. \eeq We may use this
form of $a(t)$ to find the Hubble parameter  and the acceleration
for small $\epsilon$.  We find \bea
H &=& \frac{\dot a}{a} \simeq H_0 \left[ 1 -\frac{\epsilon}{\beta}\left[\frac{B}{(\mu t+\nu)}+\frac{2C}{(\mu t+\nu)^2}\right]\right] \label{Hpert}\\
qH^2 &=&  \frac{\ddot a}{a}\simeq q_0H_0^2\left[ 1
-\frac{2\epsilon}{\beta}\left[\frac{B}{(\mu
t+\nu)}+\frac{(2\beta-3)}{(\beta-1)}\frac{C}{(\mu
t+\nu)^2}\right]\right] \label{qpert} \eea
where $H_0$ and $q_0\,H_0^2$ are the zeroth-order values of these parameters. Note
that in the large time limit ($t\rightarrow\infty$) these terms tend
to their zeroth-order values.  Hence, we see that the Gauss-Bonnet
contributions become vanishingly insignificant for late cosmological
times as one would expect.

\subsection{Exact Solution when $\eta_5/ \eta_4= \xi_4/\xi_5$}

We will now examine the equation for $p_d$ (eqn \ref{presD}) when the ratio of the coefficients of the Gauss-Bonnet terms is
equal to the ratio of the coefficients of the Einstein terms or
\beq
\frac{\eta_5}{\eta_4}=\frac{\xi_4}{\xi_5} \, .
\eeq
This equation holds for a wide variety of $n$ and $d$.
Notice that for the special case of parameters the equation for the $d$-dimensional pressure (eqn \ref{adot}) can be factored as
\beq
0=\left[\frac{\ddot a}{a} +\frac{\eta_5}{\eta_4}\frac{ \dot a^2}{a^2} \right]\left[\eta_4+\epsilon\xi_5\frac{ \dot a^2}{a^2}\right] \, .
\eeq
This equation can be solved exactly and has a solution
\beq
a(t) = (\mu t + \nu)^{\eta_4/(\eta_4+\eta_5)}
\eeq
 This solution can be used to find an expression for the energy density and effective pressure from the remaining field equations.
 As was done above, we can eliminate $a(t)$ and obtain an expression relating  $\tilde{p}$ and $\rho$.\beq\label{exEoS}
\left[1-\epsilon\frac{\sigma}{\zeta(1-w/\zeta)}(\tilde{p}-\zeta\rho)\right]^2=1+2\epsilon\sigma\rho
\eeq
where we defined the constants
\beq
\sigma=\frac{\xi_1}{\eta_1^2\beta}\;\;,\;\;\zeta=\frac{1}{\xi_1}\left(\gamma_2-\gamma_1\frac{\eta_5}{\eta_4}\right)
\eeq
with $w$ given by (\ref{EoSl0}).
This equation relating $\rho$ and $\tilde{p}$ is inherently {\it non-linear} as one might expect from an exact solution to Gauss-Bonnet FRW cosmology.
For late cosmological times, we expect a small Gauss-Bonnet contribution to the field equations (\ref{field}).
In this small $\epsilon$ regime, (\ref{exEoS}) reduces to
\beq\label{EoSs}
\left(\tilde{p}-w\rho\right)-\frac{\epsilon}{2}\frac{  \sigma}{\zeta(1-w/\zeta)}(\tilde{p}-\zeta \, \rho)^2\simeq 0
\eeq
Notice that in the $\epsilon\rightarrow 0$ limit (no Gauss-Bonnet terms), this expression yields a 4D equation of state (\ref{EoSl0}).  For a small but non-zero
$\epsilon$, we have a small correction term to the FRW equation of state. This equation can be expressed as
\beq
\tilde{p}= \rho \,w +\epsilon \rho^2 w'
\eeq
where $w' = \left( ((8 (w+3)(d^2+d-3))/(d(d+2))\right)$
for the special case of $n=2/d $.

We can also examine the regime where the Gauss-Bonnet term dominates over that of the Einstein term   for early cosmological times.
To examine this regime, we take the large $\epsilon$ limit of (\ref{exEoS}) and obtain
\beq\label{EoSl}
\left(\tilde{p}-\zeta\rho\right)-\frac{2}{\epsilon\sigma}\;\zeta(1-w/\zeta)(\tilde{p}-w\rho)\simeq 0
\eeq
Notice that in the $\epsilon\rightarrow\infty$ limit, we obtain a linear relation between the density and pressure as in (\ref{EoSl0})
but with the parameter $w$ being replaced by $\zeta$.  


\subsection{Asymptotic solution when $\lambda=constant$}
In the previous subsection, we examined Gauss-Bonnet FRW Cosmology for a vanishing cosmological constant.
 For a non-zero value of $\lambda$, the higher dimensional pressure equation is of the form
\beq
\frac{p_d}{2 \Upsilon} = 0= \eta_4 \frac{\ddot a}{a} +\eta_5 \frac{\dot a^2}{a^2} +\lambda+ \epsilon \left( \xi_4 \frac{\dot a^4}{a^4} + \xi_5 \frac{\ddot a}{a}\frac{\dot a^2}{a^2}\right)\eeq
Again, we may solve this equation perturbatively.  The zeroth-order equation yields the solution
\beq
a_0(t) = \left[A\exp{(\delta\;t)}+B\exp{(-\delta\;t)}\right]^{\beta}
\eeq
where $\delta^2=\lambda(\eta_4+\eta_5)/\eta_4^2$.  As was done in the previous subsection, we may use the remaining zeroth-order field equations to obtain an expression relating  $\tilde{p}$ and $\rho$.  We find the relation
\beq
\tilde{p}_0=w\rho_0+4\beta(1+2w)\lambda
\eeq

The first-order perturbation is of the form
\beq\label{1st}
 \ddot a_1(t) +\left(2\frac{\eta_5}{\eta_4}\frac{ \dot a_0}{a_0}\right)\dot{a}_1(t)+\left(3 \frac{\ddot a_0}{a_0}+ 2\frac{\eta_5}{\eta_4} \frac{\dot a_0^2}{a_0^2}+4\frac{\lambda}{\eta_4}\right)a_1(t)=-\frac{1}{\eta_4}\left(\xi_4 \,\frac{ \dot a_0^4}{a_0^4}+ \xi_5\frac{ \ddot a_0}{a_0}\frac{ \dot a_0^2}{a_0^2}\right) \, a_0
\eeq
Using the solution for $a_0(t)$, (\ref{1st}) becomes
\beq
\ddot a_1(t) +\left(2\frac{\eta_5}{\eta_4}\tilde{\beta}\right)\dot{a}_1(t)+\tilde{\beta}^2 a_1(t)=-\frac{\tilde{\beta}^4}{\eta_4}\left[\xi_4 - \xi_5\left(1+2\frac{\eta_5}{\eta_4}\right) \right]A\exp{(\tilde{\beta}\; t)}
\eeq
where we set $B=0$ as we are interested in an expanding universe and defined the constant $\tilde{\beta}^2=\lambda/(\eta_4+\eta_5)$.
The solution for $a_1(t)$ can be found and is
\beq
a_1(t)=a_{1, hom}(t)+a_{1, non}(t)\eeq
where the homogeneous and transient solutions are given by
\bea
a_{1, hom}(t) &=&a_0(t)\left[c_1\;e^{\tilde{\beta}(\gamma-1/\beta)\;t}+c_2\;e^{-\tilde{\beta}(\gamma+1/\beta)\;t}\right]\nonumber\\
a_{1, non}(t)&=&-\chi\;a_0(t)
\eea
where $c_1$ and $c_2$ are arbitrary constants and
\bea
\gamma^2&=& \left(\eta_5^2/\eta_4^2-1\right)\nonumber\\
\chi&=&\frac{1}{2}\frac{\eta_4\tilde{\beta}^2}{\eta_5(\eta_4+\eta_5)}\left[\xi_4 - \xi_5\left(1+2\frac{\eta_5}{\eta_4}\right) \right].
\eea
 Notice that the sign of the quantity $\gamma^2$ determines whether the scale
factor experiences oscillatory motion or exponential growth and decay.  Explicitly, exponential growth and decay occur when $\eta_5/\eta_4>1$
or $\eta_5/\eta_4<-1$.  In terms of $n$ and $d$, when
\bea
(d-1)\left(n-\frac{3}{d}\right)^2+3\frac{(d+3)}{d^2}&>&0\label{ineq1}\\
n(d-1)(2-d \, n)&>&0\label{ineq2}
\eea
Notice that (\ref{ineq1}) is trivially satisfied for all permissible values of $n$ and $d$.  However, (\ref{ineq2}) is only satisfied when
$n<2/d$; thus $n=2/d$ represents a critical point of the system for which value the scale factor behaves as if it is {\it critically damped}.  For this critical point, we find that the coefficients dramatically simplify to
$$\eta_1=\eta_3=\eta_4 = \eta_5 = - \frac{2+d}{d}\;\;,\;\;\eta_2=0$$.

As in subsection \ref{lambda0}, we may use this value of $a(t)$ to calculate the Hubble and acceleration parameters in the limit where the Gauss-Bonnet terms
are small with respect to the Einstein terms:
\bea
H=\frac{\dot a}{a} &\simeq& H_0\left[1+ \epsilon \left[(\gamma-1/\beta)c_1e^{\tilde{\beta}(\gamma-1/\beta)\;t}-
        (\gamma+1/\beta)c_2e^{-\tilde{\beta}(\gamma+1/\beta)t}\right]\right]\\
qH^2=\frac{\ddot a}{a}&\simeq& q_0H_0^2\Bigg[ 1- \epsilon \Big\{\left(1-[\gamma+(1-1/\beta)]^2\right)c_1e^{\tilde{\beta}(\gamma-1/\beta)\;t}- \nonumber\\
 && \hspace{3cm} \left(1-[\gamma-(1-1/\beta)]^2\right)c_2e^{-\tilde{\beta}(\gamma+1/\beta)t}\Big\} \Bigg]
 \eea
Note that these terms are either exponentially decaying or growing.

\section{Conclusion} \label{conclusion}
To summarize, we have found the form of the conformal factor, $ a(t)
$, within the context of an action that is the sum of Einstein and
Gauss Bonnet terms in $4+d$ dimensions with a compactifing
$d$-dimensional flat Kaluza-Klein space for a variety of cases.  By assuming that the extra dimensions compactify as
$a(t)\sim b(t)^{-n}$, we showed that restrictions on the possible form of the equation of state are imposed. Using
 pertubative methods to solve the field equations, a general form for
$a(t)$ to order $\epsilon$ was found for both a zero and a
non-zero cosmological constant . We used these values of $a(t)$ to calculate values for both the
Hubble and the acceleration parameters $\left(H(t), q(t) \right)$
and found a first order correction due to the Gauss Bonnet terms
which can play the role of additional source terms. These solutions
are consistent with having a dark energy era for reasonable values
of $n$ and $d$.  An interesting subcase is when
$\eta_5/ \eta_4= \xi_4/\xi_5$, for these values of $n$ and $d$ we
are able to solve the $d$ dimensional pressure equation exactly.
This is consistent with and generalizes the results of both Paul and Mukherjee\cite{PaulMuk} and
Mohammedi \cite{Mohammedi} . A special case happens if
$n=2/d$, when this is true then the equations simplify considerably
as detailed above and may be indicative of an underlying symmetry.

The Gauss Bonnet terms can only contribute significantly during the
early phase of the universe as one would surmise from dimensional
arguments since the Einstein terms go as $R \sim t^{-1/2}$ but
$\mathcal{G} \sim R^2 \sim t^{-1}$ etc. It is apparent that the
Gauss-Bonnet term does not contribute significantly to any Big Rip
scenario and that the current EoS with exotic matter dominating over
baryonic matter still gives rise to this singularity in the future.
A scenario where a scalar field coupled to a Gauss-Bonnet term that
avoids a Big Rip has been investigated by Nojiri, Odinstov and
Sasaki \cite{Sasaki}. However in the early universe the phantom
and/or quintessence energies need not dominate over the Gauss-Bonnet
contributions and the EoS in this regime can still be radiation
dominated.

This paper leads to several questions. What explicit mechanism might
drive cementification and on what timescale does this occur? First
how does the coupling constant $\left(\Upsilon\right)$ run as the
extra dimensions compactify? A second issue is how the addition of
these Gauss Bonnet terms would change the semi-classical states of
FRW solutions to the Wheeler-de Witt equation. The general relation
of time scales is also important since it is not uncommon to have
all of the critical compactification processes complete on the order
of a Planck time. These issues will be reported on in future papers.

 \end{document}